\documentclass[11pt,preprint]{aastex}
\usepackage{color}
\usepackage{natbib}

\shorttitle{PSR J1903+0327 : A Unique Milli-Second Pulsar with a Main-Sequence Companion Star }
\shortauthors{Khargharia et al.}

\begin{document}


\title{PSR J1903+0327 : A Unique Milli-Second Pulsar with a Main-Sequence Companion Star }


\author{Juthika Khargharia, John T. Stocke, Cynthia S. Froning}
\affil{Department of Astrophysical and Planetary Sciences,}
\affil{Center for Astrophysics and Space Astronomy}
\affil{University of Colorado, 391, UCB \\ Boulder, CO, 80309}
\email{juthika.khargharia@colorado.edu, john.stocke@colorado.edu, cynthia.froning@colorado.edu}

\author{Achamveedu  Gopakumar}
\affil{The Department of Astronomy and Astrophysics}
\affil{Tata Institute of Fundamental Research \\ Mumbai, India, 400005}
\email{gopu.tifr@gmail.com}

\and

\author{Bhal Chandra Joshi} 
\affil{National Center for Radio Astrophysics}
\affil{ P.~O.~Bag \# 3 \\ Ganeshkhind, Pune, India, 411007} 
\email{bcj@ncra.tifr.res.in}




\begin{abstract}

PSR J1903+0327 is a mili-second pulsar with a mass of 1.67$M_{\odot}$ in a highly eccentric orbit ($e=0.44$) around a main-sequence star. This unique system cannot be reconciled with current observations where milli-second pulsars are generally seen to orbit white dwarfs in almost exactly circular orbits. Current theoretical models of binary and stellar formation and evolution cannot explain the high eccentricity of this system either.  In this work, we present three new epochs of optical spectroscopy for the companion to PSR J1903+0327, obtained to confirm the association of the main-sequence star with the pulsar. These 3 new epochs together with the 2 previous ones obtained by \citet{freire} firmly establish the high eccentricity of the companion's orbit as predicted by pulsar timing. Using all five epochs of optical data, we have provided an independent estimate of the mass-ratio, $R=1.56\pm0.15$ as well as the systemic radial velocity of the binary, $\gamma= 42.1\pm2.5$ km s$^{-1}$. We constrain the spectral type of the pulsar companion to lie between F5 V -- G0 V (a slightly earlier type than suggested previously) by measuring the equivalent widths of two of the three Ca-triplet lines (8498 \AA, \  8542 \AA ) and the Paschen line at 8598 \AA \ (P14); we also broadly constrain the metallicity of the companion. Additionally, we have placed a somewhat better limit on the rotational velocity of the pulsar companion of $v_{rot}\sin{i} \leq66$ km $s^{-1}$ which is still not sufficient to allow a test of general relativity using this system. 
\end{abstract}





\newpage

\section{Introduction}

Binary pulsars are excellent systems for understanding the nature of binary star evolution and they also act as laboratories for tests of extreme physics. The milli-second pulsar (MSP), J1903+0327, was first discovered in the Arecibo L-band Feed Array pulsar survey \citep{cordes}. It was soon found by \citet{champion} that the MSP has a spin period of 2.5 ms and is in a highly eccentric orbit (${\it{e}} =0.44$) around what appeared to be a solar-mass companion. It is atypical of MSPs to exhibit such high eccentricity or to have a main-sequence (MS) companion. Current models of stellar and binary evolution do not predict such a scenario either \citep{stairs04}. This is the first and only MSP in the Galactic disk that exhibits such orbital attributes, and therein lies its uniqueness. To search for the pulsar companion, \citet{champion} obtained near-infrared J-, H- and K$_{s}$ images of the pulsar field with the Gemini-North telescope and found a single star at the position of the pulsar with J=19.22, H=18.41 and K$_{s}$=18.03 magnitudes. They determined, from the density of stars in the field, that the probability of finding a star at the position of the pulsar is $<3\%$. However, it was not known whether this star was a chance encounter or a distant third member in a hierarchical triple system.  To confirm this association or lack thereof, \citet{freire}; F2011 hereafter, obtained optical spectroscopy of the possible pulsar counterpart using the FOcal Reducer and low dispersion Spectrograph (FORS2) on the European Southern Observatory's (ESO) Very Large Telescope (VLT). They determined the spectral type of the companion to lie between an early to mid-G dwarf and from radial velocity changes, they positively confirmed the MS star's close-in association with the pulsar. It is typically seen that MSPs with spin-periods $<10$ ms have white dwarfs (WD) as their companions and exhibit very circular orbits with eccentricities, ${\it{e}} < .001$ \citep{phinney}. To explain this inconsistency in PSR J1903+0327, \citet{champion} proposed the possibility of a triple system where the pulsar is orbiting an unseen massive white dwarf (WD) of 0.9--1.1 $M_\odot$ while the MS star is in a much wider orbit around the MSP-WD binary and provides the eccentricity to the inner binary via the Kozai mechanism, which causes periodic exchange between inclination and eccentricity of the orbit \citep{kozai}. Recently, \citet{zwart} have questioned the plausibility of the existence of such a MSP--WD--MS triple system and have pointed out several shortcomings associated with such a proposition. It should be mentioned that triple systems with eccentric orbits have been observed in the past. For example, the soft X-ray transient, 4U 2129+47 (V1727 Cyg) is a low mass X-ray binary (LMXB) with an orbital period of 5.24 hr and a third companion (F-type MS star) in an eccentric orbit of 175 days \citep{garcia, bothwell, lin}; similarly the cataclysmic variable EC 19314-5915 has an orbital period of 4.85 hr with a G8 V star as the outer companion \citep{buckley}.  An alternative scenario proposed by \citet{champion} was the possibility of an exchange interaction in a globular cluster that eventually resulted in the MSP orbiting the MS star as its present companion, but the new observations of F2011 have shown this to be highly unlikely.  Other formation and evolution mechanisms discussed in F2011 are summarized in Section 4.2. In the same section, we also discuss the results obtained from the numerical simulations by \citet{zwart} to test the formation scenario of PSR J1903+0327 proposed by F2011. \\

Based on the observations of \citet{champion} and F2011, we appear to have the unique case of a MSP in orbit around a MS star in a highly eccentric orbit in the Galactic disk. This can prove to be of immense importance in the field of general relativity. Pulsar timing of this unique system can test alternative theories of gravity and probe the equation of state (EOS) of super-dense matter. Accurate pulsar masses already set interesting limits for the EOS of neutron star (NS) material \citep{freire09}, ruling out a few softer models. Also, the rate of orbital decay of this system will constrain various scalar-tensor theories of gravity because the members of this binary system have vastly different self-gravities, ( $0.2$ for NSs and $10^{-6.0}$ for MS stars). An accurate measurement of the orbital decay of the pulsar companion accompanied by the confirmation that it is a slowly rotating solar mass MS star can constrain a significant portion of parameter space for general relativity versus scalar tensor theory. So far, the best candidate for constraining these dipole contributions consists of a 394 milli-second pulsar orbiting a WD with comparable mass in a moderately eccentric orbit \citep{bhat}. In F2011, the rotational velocity of the pulsar companion could not be very tightly constrained ( $< 140$ km $s^{-1}$, 3-$\sigma$). An accurate estimate of the rotational velocity of the pulsar companion is important to constrain the classical contributions to the measured periastron advance of the MSP. The dominant classical contributions arise from the rotationally-induced quadrupole moment of the star and this classical spin-orbit coupling can account for 10$\%$ of the measured periastron advance for a rotational period of a few days \citep{wex}. \\

In this paper, we present three new epochs of optical spectroscopy of the pulsar companion obtained with the Gemini Multi-Object Spectrograph (GMOS) at Gemini-N. In Section 2 we describe the observations and data reduction procedures. In Section 3, we present a detailed analysis to confirm the results of F2011 as to the association of the pulsar companion with the MS star, determine the spectral type of the pulsar companion as accurately as possible using a combination of F2011 data along with our new spectra and constrain the rotational velocity of the pulsar companion more precisely.  Section 4 discusses the various formation and evolution scenarios possible for this system. The result obtained on the rotational speed of the pulsar companion has several implications in the field of general relativity which are also presented in this section. Finally, we summarize the conclusions in Section 5.

\section{Observations and data reduction}

Long slit (1.0$^\prime$$^\prime$ slit-width), moderate resolution (R=2200) spectroscopy of the pulsar companion in the 7500--9600 \AA \ range was performed with GMOS on Gemini North on three different observing epochs separated by  20 -- 40 days. The observations took place on Aug 19, Sep 06 and Oct 05, 2010 using the EEV detector with five 30 minute individual exposures (total of 2.5 hours on-source integration time). Wavelength calibration exposures were obtained before and after each target exposure, which were averaged to determine a wavelength solution for that particular exposure. Wavelength solution accuracy was further checked using the sky lines in the spectrum. Additionally, a bright slowly rotating (11 km$s^{-1}$) F-type star was observed before and after the target observations which served as our rotational standard star \citep{wolff}.  The detector was read out with no binning, giving a resolution of 3.4 \AA\ sampled at 0.34 \AA\ pix$^{-1}$. The Gemini IRAF package for GMOS data reduction was used to reduce the data which involved the processes of bias subtraction, flat-fielding, sky-subtraction and finally, spectral extraction. No flux calibration was performed on the data. F2011 points out that a bright star located 2.3$^\prime$$^\prime$ from the pulsar counterpart contributes partly to the counts from the latter, complicating its extraction in their spectrum. Hence they used the optimal spectral extraction technique by \citet{hynes} to obtain the spectrum of the pulsar companion. In Fig. 1, we show the position of the target compared to the bright source located 2.3$^\prime$$^\prime$ away. Upon fitting Moffat profiles to both the target and the bright object, we notice that in the better seeing conditions of the GMOS observations, the contamination caused by the bright object to the target is negligible. Additionally, we selected a narrow spatial window (0.5$^\prime$$^\prime$ -- 0.6$^\prime$$^\prime$) for spectral extraction of the target, thus further reducing any contamination from the bright source. Our final extracted spectra still has residuals from night sky lines. Furthermore, the wavelength range from 8600 --- 8700 \AA \ is affected by the broad feature in the airglow spectrum which is contributed by a blend of R- and P- branches of $O_{2} (0 -1)$ \citep{broadfoot}. This, along with residual night sky lines, resulted in our inability to recover the \ion{Ca}{2} 8662 \AA \ feature in the spectrum of the pulsar companion. In the remainder of the paper, we will focus on the wavelength region 8400 -- 8600 \AA \ where most of the prominent features reside and where our sky subtraction was best. The features in this region were identified using the line list from \citet{cenarro1} and the night sky lines were identified from \citet{osterbrock}.

\section{Analysis}

In Section 3.1, we confirm the results of F2011 as to the association of the MS star to the pulsar by comparing the radial velocities obtained by us in three different epochs against the radial velocity of the pulsar companion obtained by F2011. Second, in Section 3.2, we constrain the spectral type of the pulsar companion by employing equivalent widths of some of the prominent absorption line features in our spectrum and comparing them against values from the literature. 
In Section 3.3, we have constrained the rotational velocity of the pulsar companion using the rotational standard star that was observed for this purpose. \\

\subsection{Radial Velocity of the pulsar companion} 

To compute the radial velocities at each of the three epochs, we first obtained the wavelength shift in the strongest \ion{Ca}{2} triplet feature at 8542 \AA. The accuracy of the wavelength calibration was established by comparing the position of night sky lines in the spectrum of the pulsar companion against the values from \citet{osterbrock}. From this comparison, we found zero point offsets in the wavelength calibration  of -0.31 \AA, \  -0.25 \AA \ and +0.15 \AA \  for the three observation epochs respectively, which were then taken in account while computing the final radial velocities. After shifting the individual spectra by their respective wavelength offsets, we created an averaged spectrum of the pulsar companion shown in Fig. 2. The calcium 8542 \AA \  and the 8498 \AA \ features are seen in the spectrum. The absorption feature at 8598 \AA \  is  at the location of the P14 line in the Paschen series. These features are present in the spectrum at each epoch. There are also possible detections of \ion{Fe}{1} at 8514.1 \AA \ and 8688 \AA. However, considering the modest S/N ($\sim$ 7 -- 10) of the spectrum, we have decided to restrict our analysis to the lines :  \ion{Ca}{2} ($\lambda$8498, $\lambda$8542) and  P14($\lambda$8598). The deviation of the data points from a straight-line fit to various regions of continuum was used to assign the error on the normalized spectrum.  In Fig. 2, the location of sky line residuals have been replaced with similar continuum regions from the GMOS spectrum of the rotational standard. These locations are marked with vertical dashed lines in Fig 2. To refine our estimates for the radial velocities, we cross-correlated the averaged spectrum with the individual unshifted ones at each epoch while correcting for the wavelength calibration accuracy. The cross-correlation function in each case was fitted with a Gaussian and the error was obtained by the uncertainty in the determination of the line centroid. This error was then propagated in the estimation of the radial velocites. As a result, we obtained radial velocities of $V_{1}= 39.1 \pm 4.5$ km $s^{-1}$, $V_{2}= 25.2 \pm 5.4$ km $s^{-1}$ and $V_{3}=17.8 \pm 6.4$ km $s^{-1}$ with respect to the solar system barycenter at the three different epochs, respectively. Using the radial velocities obtained by us and F2011 at five different epochs, we can estimate the systemic radial velocity, $\gamma$ of the binary, which cannot be derived from radio timing. Towards that end, we minimized the chi-square, $\chi^2_{\nu}$, between the observed radial velocities and the corresponding predicted values using the parameters determined from radio timing, while varying $\gamma$. By using all five data points, we have refined the estimate for the systemic velocity given in F2011 to $\gamma = 42.1\pm2.5$ km s$^{-1}$ ($\chi_{\nu}=0.65,\ 1\sigma$), a value which supports the suggestion of F2011 that PSR J1903+0327 did not originate in a dense stellar cluster. Using this value of the systemic velocity, we next minimized the chi-square as before while varying the mass-ratio alone. From this, we obtained an independent estimate of the mass-ratio, R, and refined the value given in F2011 to $R=1.56\pm0.15$ ($\chi_{\nu}=0.50$,\ 1$\sigma$). This estimate is in agreement with the result derived from radio timing. We did a similar analysis to constrain the eccentricity, $e$, of the binary using five epochs of optical observations and found this number to lie between $0.42 \leq e \leq 0.58$ ($\chi_{\nu}=0.40$,\ 1$\sigma$), thereby confirming the high eccentricity of the binary. Further optical/infrared observations of the pulsar companion could be useful in placing tighter constraints on the parameters discussed above. Fig. 3 shows the predicted radial velocity curve of the pulsar companion (solid line), based on the orbital parameters determined from pulsar timing and the systemic radial velocity obtained from five epochs of optical data. The dotted-dashed lines in the same figure shows the radial velocity curve of the pulsar companion for an eccentricity of $e=0.50$, mass-ratio of $R=1.62$ and systemic velocity of $\gamma=42.1$ km s$^{-1}$ (dotted-dashed line). Our observations confirm the result of F2011 that the optical counterpart is indeed the close-in companion to the pulsar, orbiting it in a 95-day period. With five epochs of radial velocities in-hand and in agreement with the predicted reflex motion of the pulsar companion, the highly elliptical orbit of this system is confirmed as well.

\subsection{Spectral type of the pulsar companion}

Significant work has been done in the past to use equivalent widths (EWs) of the calcium triplet as well as the Paschen lines as indicators of temperature, metallicity, and luminosity of a star. This is the method we have employed in determining the spectral type of the pulsar companion. To this end, we have followed the works of \citet{mallik94}, \citet{ginestet}, \citet{jones}, \citet{idiart} and \citet{diaz}. The spectral features along with the line windows used for calculation of EWs and our results are shown in Tables 1 and 2. EWs were calculated from the  the averaged, normalized spectrum and errors on EW values were obtained by choosing different continuum points for normalization ($\pm 1 \sigma$). Since the \ion{Ca}{2} line at 8662 \AA \ was not recovered in our spectrum, we have used the EW of  this feature from the normalized spectrum of F2011 wherever applicable (see Table 2). Furthermore, we have reanalyzed the data of F2011 using the two well-detected \ion{Ca}{2} features at 8542 \AA \ and 8662 \AA \  shown in Table 3. This was done to ascertain if tighter constraints could be placed on the spectral type of the pulsar companion from their data. It should be noted that the error in the EWs obtained from their already normalized spectrum is based on the error in the Gaussian fit to the feature rather than on continuum placement.

A summary of our conclusions from various references are shown in Table 4. The sum of EWs of the calcium triplet is a robust indication of the luminosity class and following all the references cited above, the pulsar companion clearly belongs to a luminosity class 'V'. It is a determination of the temperature that poses a greater challenge. Using the EWs of the calcium triplet alone does not constrain the spectral type any tighter than between F5 -- G5 ( see Table 4). On the other hand, from \citet{ginestet}, the non zero EW of the P14 line restricts the spectral type to G0 or earlier. Our confidence in the P14 detection is supported by the fact that we observe this feature in all three epochs of data with the same wavelength shift as the calcium lines at 8498 \AA \ and 8542 \AA. Taking this into account, we conclude that the spectral type of the pulsar companion lies between F5  V -- G0 V. The last column in Table 4 summarizes the metallicity obtained for the pulsar companion from each of the references. While indicative, we do not attempt to draw any significant conclusion from it since the metallicity results are limited by the uncertainty in the spectral type.

F2011 determined the spectral type of the pulsar companion by comparing their averaged spectrum with synthetic spectra and obtained a temperature of $T_{eff} = 5825 \pm 200 K$ and a surface gravity of $\geq$ 4.0 cm s$^{-1}$. Their analysis points to a star roughly of spectral type G0 -- G5 V \citep{ali}.  However, from an equivalent width analysis of the \ion{Ca}{2} features at 8542 \AA \ and 8662 \AA \  obtained from their data, we can conclude only that the star has a spectral type between F5 V -- G5 V. But we also note that they appeared to have detected P14 as we do although at the low signal-to-noise based on our own measurement (see Table 4), constraining the spectral type to G0 or earlier. There is a slight mismatch between our conclusion and F2011 regarding the spectral type of the pulsar companion ( F5 V -- G0 V vs G0 V -- G5 V) . However, the detection of Paschen P14 line in all three epochs of our data and that of F2011 requires a spectral type G0 or earlier. Taking both studies into consideration, G0 V appears to best describe the spectral type of the pulsar companion. This also matches the prediction of the companion's mass of 1.667M$_{\odot}$ from pulsar timing (F2011). 

\subsection{Rotational velocity of the companion star} 

We observed a bright rotational standard star (HD 176095) with GMOS at each epoch before and after our target observations. The rotational standard star is an isolated, slowly rotating (11 km$s^{-1}$) late F type star \citep{wolff}. 
Fig. 4 shows the spectrum of the pulsar counterpart plotted over the spectrum of the rotational standard (in red).
At a spectral resolution of 136 km s$^{-1}$, the effect of broadening from a star rotating at 11 km s$^{-1}$ cannot be detected. Further, the calcium lines in solar-like stars are often broadened by collisional or pressure effects in the star's atmosphere \citep{smith}. From the spectrum of the pulsar companion, we find that the FWHM of the \ion{Fe}{1} feature at 8515 \AA, which is insensitive to pressure broadening, is similar to the FWHM of an arc lamp line, indicating that the pulsar companion is not rotating at a speed higher than the instrumental resolution. Therefore, an upper limit on the rotational velocity of the pulsar companion can be set by finding the minimum broadening that is detectable at the resolution of our instrument using cross-correlation techniques. From \citet{tonry} and \citet{bailer}, the measured width (FWHM) of the cross-correlation function (CCF) can be expressed as 

\begin{equation}
\sigma^2_{meas} = \sigma^2_{rot} + \sigma^2_{nat} + 2\sigma^2_{inst}
\end{equation}

\noindent where $\sigma_{rot}$ is the rotational broadening, $\sigma_{nat}$ is the intrinsic broadening and $\sigma_{inst}$ is the instrumental broadening. Following the argument made in \citet{bailer}, $\sigma_{rot}$ can only be detected if it exceeds $\sqrt(2) \sigma_{inst}$, assuming that $\sigma_{nat}$ is negligible. Also, the full width of a rotational profile is contributed equally by both the approaching and the receding limb of the star and hence corresponds to twice the rotational velocity. Theoretically, we should be able to detect a minimum rotational velocity of $v_{rot} \sin{i} =\sqrt(2)(\sigma_{inst}/ 2)$ \citep{bailer}. 

To test that, we obtained the autocorrelation function (ACF) of the rotational standard star separately in the wavelength ranges 8000 -- 8450 \AA \ and 8670 -- 8800 \AA \ . These regions mostly comprise the \ion{Fe}{1}, \ion{Mg}{1} and \ion{Ti}{1} features while excluding the pressure sensitive calcium features. The average of the FWHMs of these two wavelength regions on either side of the calcium triplet was used as a measure of $\sigma_{meas}$. We then artificially broadened the spectrum of the rotational standard in steps of 3 km s$^{-1}$ and cross-correlated the broadened spectrum with the original un-broadened one until a measurable difference was noticed in the CCF. From Eq. 1, $\sigma^2_{meas} = 2\sigma^2_{inst}$ (for the ACF) and $\sigma^2_{meas} = \sigma^2_{rot} + 2 \sigma^2_{inst}$ (for the CCF) and subtraction of one from the other gives $\sigma_{rot}$. We determined FWHMs of the ACF and the CCFs by fitting them with Gaussian profiles. A measurable difference in the CCF was obtained when the rotational standard star spectrum was broadened by $\sigma_{rot}$ = 132 km s$^{-1}$. Hence, the minimum detectable $v_{rot} \sin{i}$ at our instrumental resolution is 66 km s$^{-1}$. This value is close to the theoretical minimum detectable rotational speed of 96 km s$^{-1}$ (=$136 \times \sqrt{2}$ /2), obtained from the argument in \citet{bailer}. Therefore, these observations constrain the pulsar companion's rotational speed to $\leq$ 66 km s$^{-1}$. Our constraint on the rotational velocity of the pulsar companion is a factor of $\sim$ 2 less than the value quoted in F2011.

\section{Discussion}

From our analysis, we have confirmed that the the observed star in the vicinity of the MSP is indeed its companion orbiting it in a highly elliptical 95-day orbit as predicted from pulsar timing. From the EWs of the calcium triplet and the P14 feature, we constrain the spectral type of the star to lie between F5 -- G0 V. Additionally, we have proven that the companion is rotating slowly with an upper limit of $\leq$66 km s$^{-1}$. In this section, we will discuss the implication of our result in two areas : constraining evolution scenarios of this unique system and using this system as an astrophysical laboratory for tests of general relativity.  
 
\subsection{Implications for General Relativity}

The present analysis should allow us to constrain the classical spin-orbit contribution to the measured rate for the periastron advance: $\dot \omega_{o} = 86.38 \pm 0.08$ arcsecond century$^{-1}$. This is because the above contribution arises from the rotationally induced quadrupolar moment of the main-sequence companion in our binary pulsar and in what follows we provide a few estimates for $\dot \omega_{{SO}}$. Using equation (79) in \citet{wex}, we obtain a maximum value for the apsidal motion due to classical spin-orbit coupling to be $\dot \omega_{SO} \sim 7.6 \times 10^{4} \times J_2$ arcsecond century$^{-1}$, where $J_2$ is the quadrupolar moment of companion star that we observed. If we let $J_2$ take a value close to that for the Sun ( $J_{2\, \odot } \sim 1.7 \times 10^{-7} $), we  get $\dot \omega_{SO} \sim 0.013 $ arcsecond century$^{-1}$ and this is roughly an order of magnitude smaller than the measurement uncertainty in $\dot \omega_{o}$(F2011). However, the fact that $J_2$ is proportional to the square of the angular velocity of the star's proper rotation implies that we may use $J_{2} \sim J_{2\, \odot } \times v_{{rot}}^2/ v_{\odot}^2$, where $ v_{rot} $ stands for the rotational velocity of the companion and we let $v_{\odot} \sim 2 $ km$\,{s}^{-1}$. Therefore, we can provide few constraints for  $\dot \omega_{SO}$ contributions to $\dot \omega_{o}$.

If we let the spin axis of the star be parallel to the orbital angular momentum of the binary, we have $v_{{rot}} \sim$ 66 km s$^{-1}$ which leads to $\dot \omega_{{SO}}\sim 14 $ arcsecond century$^{-1}$. This is even substantially higher than the $2.3$ arcsecond century$^{-1}$ upper bound on the spin-orbit contributions to the apsidal motion that was obtained by treating general relativity to be the correct theory and using the minimum total mass compatible with the measurements of three relativistic observables (F2011). Interestingly, the fact that we have restricted the spectral type of the companion between an F5V -- G0 V, implies that $v_{{rot}}$ may be in the range $ 12-25$ km ${s}^{-1}$ \citep{tassoul}. These rotational velocity bounds lead to two estimates for  $\dot \omega_{{SO}}$: $ 0.5$ and $ 2$ arcsecond century$^{-1}$ respectively. These are also higher than the present  measurement error in  $\dot\omega_{{o}}$ indicating that it will be rather difficult to perform a test for general relativity using the available measurements of the three post-Keplerian parameters (F2011).

Our 8 meter spectra were taken under good to excellent conditions and yet have a relatively modest signal to noise at a resolution of R=2200 which is still $\sim$ 7--8 times poorer than what we need to achieve a constraint of less than 10 km s$^{-1}$ on $v_{rot} \sin{i}$. The lines which are unaffected by pressure broadening in the star's atmosphere are too weak to be detected clearly at the current resolution and S/N and would require much better signal-to-noise to be confidently detected. Therefore, employing the agreement of the three different post-Keplerian parameters to perform a test of general relativity is not plausible at the current time.

\subsection{Constraining evolution scenarios for the pulsar system}

The detection of a MS star as the companion to the pulsar and the confirmation of a highly elliptical orbit raises various questions on the origin of this system while eliminating some other hypotheses.  For example, \citet{champion} considered the possibility of a triple system where the pulsar is orbiting an unseen white dwarf (0.9 -- 1.1$M_\odot$) in a 95-day orbital period while the MS star happens to be in a longer period that drives the eccentricity of the system by the Kozai mechanism \citep{kozai}. Since we have established that the MS star orbits the pulsar in a 95-day orbital period, the triple system hypothesis can be ruled out. Additionally, \citet{zwart} have pointed out that the hypothesis of a MSP--WD--MS triple system is precluded by the fact that the observed change in the eccentricity of this system is three orders of magnitude smaller than that predicted via the Kozai mechanism \citep{gopakumar} .

\citet{champion} also discuss the possibility of an exchange interaction in a dense stellar environment that might have brought the MS star to orbit the pulsar in a 95-day period. In F2011, the probability of this was found to be negligible. One of the parameters used in their simulation is the systemic radial velocity, $\gamma$, of the binary. With five epochs of radial velocity data, we have now refined this number to be $\gamma=42.1\pm2.5$ km s$^{-1}$. This is consistent with the value obtained by F2011 thereby supporting their suggestion that the system did not form in a dense stellar environment.

F2011 also find that the pulsar has been recycled or spun-up to milli-second speeds (due to mass-transfer from a past, accreting X-ray binary phase) but show that the current MS star is unlikely to have played the role of the donor star responsible for its recycling. To answer the question of how the pulsar was spun-up and what happened to the donor star, F2011 investigated another triple system scenario. They considered a triple system wherein the outer companion (the present MS star) started out in a much wider orbit around the inner binary, which consisted of two more massive MS stars in a shorter orbital period. The more massive star in the inner binary evolves into a red supergiant phase and then engulfs the companion into a common envelope phase. As the orbit of the inner binary decays, it will release orbital energy sufficient enough to leave a much closer binary consisting of a He-core and an almost unaffected MS star. Meanwhile, if the outer star in the triple system is close enough, it could be engulfed by the expanding envelope of the inner binary as well, which will cause it to spiral in and result in a much closer triple system. Once the He-core explodes in a supernova, we are left with an inner Low-Mass/Intermediate-Mass X-ray Binary (LMXB/IMXB) and an outer MS star. The NS is then spun up to milli-second periods by the donor star, which has filled its Roche-lobe by that time. As mass transfer continues from the donor to the NS, the orbit of the inner binary in the triple will widen to some extent and will end in  a chaotic three-body interaction. This will result in expulsion of the less massive star from the system, eventually leaving the MSP in a tighter and eccentric orbit around the MS star (the initial outer body in the triple system). F2011 have also discussed in less detail several alternative scenarios for the formation and evolution of the system. All of these scenarios start out with the system as part of a triple and then invoke different dynamics that transpires after the common envelope phase, leading to the system as we see it now. However, thorough numerical simulations are necessary to provide more insight into the formation and evolution of these systems in a non-deterministic way.

Recently, \citet{zwart} performed a detailed numerical analysis to test the formation scenario presented in F2011. They start out with a stable LMXB ($P_{orb} < 1.0 d$) with a MS star in a wide and possibly eccentric outer orbit. They have shown from their simulations that as mass transfer starts in the inner binary and as the orbit of the inner binary widens, the system becomes unstable. Depending on the exact configuration of the system at this point, each one of the three components in this triple system could be ejected. Specifically, they found that a stable hierarchical triple consisting of an inner binary with component masses of  9--13 $M_{\odot}$ and 0.8 -- 2.0 $M_{\odot}$, separated by $\geq 200 R_{\odot}$ and an outer MS star with mass $<$ 2$ M_{\odot}$ and having a semi-major axis $>$ 560$ R_{\odot}$, could eventually result in the formation of a system like PSR J1903+0327. Even though the chances of passing through the various chain of events were found to be low, the result matches with what we see in PSR J1903+0327. \citet{zwart} were also able to describe an evolution scenario for the soft X-ray transient 4U 2129+47 (V1727 Cyg) and the cataclysmic variable, EC 19314-5915, we noted before in Section 1. Additionally, they concluded that systems like PSR J1903+0327 have a birthrate less than 3\% compared to the Galactic LMXB's and 10 times less than all MSPs. Thus the probability for this formation scenario happening, is not large, but it is also not nil. Their model predicts tens to hundreds of systems similar to PSR J1903+0327 existing in our Galaxy and raises the question of why no system like this has been found prior to PSR J1903+0327. 

In another independent work, \citet{liu} consider that PSR J1903+0327 started out as a binary system but during the supernova (SN) explosion that produced the pulsar, a fall-back disk was formed around the newborn NS. Accretion started from the disk onto the NS, which spun it up to milli-second speeds. Meanwhile the high eccentricity is considered to be a consequence of the SN explosion that produced the NS. \citet{liu} mention that if this system was undisrupted, it could maintain this eccentricity for $\geq 10^{10}$ yr. One of the uncertainties mentioned in their work is whether the fall-back material would have enough angular momentum to produce the disk. This calls for an unusual SN-fallback history to explain the very short period (2.5 ms) of PSR J1903+0327. In order for the fallback disk scenario to work, it requires the companion of the pulsar to be $\leq (1-2) \times 10^{9}$ yr. From our work as well as from F2011, we find the MS star to be older than what is required for the fallback disk scenario to work. 

\section{Conclusion} 

By obtaining new spectroscopy with GMOS at Gemini-N, we have confirmed the results of F2011 which positively identified a MS star as the close-in pulsar companion. Radial velocities are now in-hand for 5 epochs and the predicted radial velocity curve obtained from pulsar timing is confirmed. The highly eccentric orbit is confirmed as well. Additionally, we obtained estimates for the mass-ratio of $R=1.56\pm0.15$ and the systemic radial velocity of the binary of $\gamma=42.1\pm2.5$ km s$^{-1}$. These refined estimates are important in the context of explaining the origin of this system. We have determined the spectral type of the pulsar companion to lie quite close to G0 V, from an analysis with our data and a reanalysis of F2011 data. Also, we have constrained the rotational velocity of the pulsar companion as $<$ 66 km$s^{-1}$. With the present estimate of the $v_{rot}\sin{i}$, it will be difficult to perform a test for general relativity. Radio pulsar timing can lead to better constraints on General Relativity in comparison to scalar tensor theories and should be pursued over the next few years. We have also noted that the evolutionary scenario possible for the formation and evolution of this unique pulsar system is best described by the recent numerical simulations of \citet{zwart}. Starting with a stable hierarchical triple system consisting of an inner binary with component masses of  9--13 $M_{\odot}$ and 0.8 -- 2.0 $M_{\odot}$, separated by $\geq 200 R_{\odot}$ and an outer main sequence star with mass $<$ 2$ M_{\odot}$ and having a semi-major axis $>$ 560$ R_{\odot}$, numerical simulations were able to replicate the formation and evolution of a system like PSR J1903+0327.




\acknowledgments{We would like to thank the Gemini staff for conducting the spectroscopic observations. We would also like to thank Cees G. Bassa for kindly providing the data for the spectrum of the pulsar companion from F2011. J. T. Stocke, A. Gopakumar and B. C. Joshi thank the Radio Astronomy Center (RAC) of the Tata Institute of Fundamental Research (TIFR) at Ootacamund, India for hospitality during a portion of this work. Additionally, J. T. S. thanks the National Center for Radio Astrophysics (NCRA) at Pune, India for hospitality during the initial stages of this work. He would also like to acknowledge the support of NSF grant AST-0707480 and an NRAO travel grant.}

Facilities: \facility{Gemini(GMOS-N)}

\clearpage
\begin{deluxetable}{cccc}
\tabletypesize{\scriptsize}
\tablecaption{Equivalent Width Analysis}
\tablewidth{0pt}
\tablehead{
\colhead{Feature} & \colhead{Integration limits ($\AA$)} & \colhead{Equivalent width($\AA$)}}
\startdata
${\bf CaT}$ ${8498}$  & 8495 -- 8501 & 0.61 $\pm$ 0.27\\
${\bf  CaT}$ ${8542}$ & 8535 -- 8550 & 1.70 $\pm$ 0.74\\
\enddata
\tablecomments{Table of EWs for comparison with the work by \citet{mallik94}}
\end{deluxetable}

\begin{deluxetable}{cccc}
\tabletypesize{\scriptsize}
\tablecaption{Equivalent Width Analysis}
\tablewidth{0pt}
\tablehead{
\colhead{Feature} & \colhead{Integration limits ($\AA$)} & \colhead{Equivalent width($\AA$)}}
\startdata
${\bf  CaT }$ ${8498}$  & 8488 -- 8510 &\ 1.17 $\pm$ 0.50\\
${\bf  CaT}$ ${8542}$ & 8530 -- 8556 &\ 2.32 $\pm$ 0.70\\
${\bf  CaT}$ ${8662}$ & 8652 -- 8672 &\  $\hspace{1mm}$ 1.36 $\pm$ 0.15$^{\tablenotemark{a}}$ \\
${\bf  P14}$ ${8598}$ & 8585 -- 8605 &\ .70 $\pm$ .50\\
\enddata
\tablecomments{Table of EWs for comparison with the work by \citet{ginestet}, \citet{jones}, \citet{idiart} and \citet{garcia_vargas}}
\tablenotetext{a}{Data for this absorption feature was used from \citet{freire}}
\end{deluxetable}


\begin{deluxetable}{cccc}
\tabletypesize{\scriptsize}
\tablecaption{Equivalent Width Analysis}
\tablewidth{0pt}
\tablehead{
\colhead{Feature} & \colhead{Integration limits ($\AA$)} & \colhead{Equivalent width($\AA$)}}
\startdata
${\bf  CaT}$ ${8542}$ & 8527 -- 8557 &\ 2.40 $\pm$ 0.17\\
${\bf  CaT}$ ${8662}$ & 8647 -- 8677 &\ 1.32 $\pm$ 0.15\\
${\bf  P14}$ ${8598}$ & 8585 -- 8605 &\ .11 $\pm$ .07\\
\enddata
\tablecomments{Table of EWs for comparison with the work by \citet{diaz}. The line window for P14 is used from Table 2. All other line windows are from \citet{diaz}. The data for computation of the above EWs is obtained from Figure 2 of \citet{freire}}
\end{deluxetable}


\begin{deluxetable}{ccccc}
\tabletypesize{\scriptsize}
\tablecaption{Summary of Results}
\tablewidth{0pt}
\tablehead{
\colhead{Reference} & \colhead{Feature used} & \colhead{Luminosity Class} & \colhead{$[Fe/H]$}& \colhead{Spectral Type}}
\startdata
\citet{mallik94} & CaT($\lambda$8498+$\lambda$8542) & IV/V & -0.3 $< [Fe/H]<$0.2 & F5 -- G5  \\
\citet{ginestet} &\ion{Ca}{2}($\lambda$8542), \ion{Ca}{2}($\lambda$8542), P14 ($\lambda$8498) & ---- & ---- & F2 -- G0  \\
\citet{ginestet} &CaT($\lambda$8498+$\lambda$8542+$\lambda$8662) & V & ---- & -----$^{\tablenotemark{a}}$  \\
\citet{jones} &CaT($\lambda$8498+$\lambda$8542+$\lambda$8662) & V & ---- & $\hspace{1mm}$ F5 -- G7$^{\tablenotemark{a}}$  \\
\citet{idiart} &CaT($\lambda$8498+$\lambda$8542+$\lambda$8662) & --- & -0.6$\leq[Fe/H] \leq$0.2 & ---$^{\tablenotemark{a}}$  \\
\citet{diaz} &CaT($\lambda$8542+$\lambda$8662) & V &  -0.4 $\leq [Fe/H] < $0.2 & $\hspace{1mm}$ F5 -- G5$^{\tablenotemark{b}}$  \\

\enddata
\tablecomments{Summary of results using the CaT features from our data and F2011}
\tablenotetext{a}{$\lambda$8662 feature was used from F2011}
\tablenotetext{b}{both features were used from the data of F2011}
\end{deluxetable}






\clearpage

\begin{figure}
\figurenum{1}
\epsscale{1.0}
\plotone{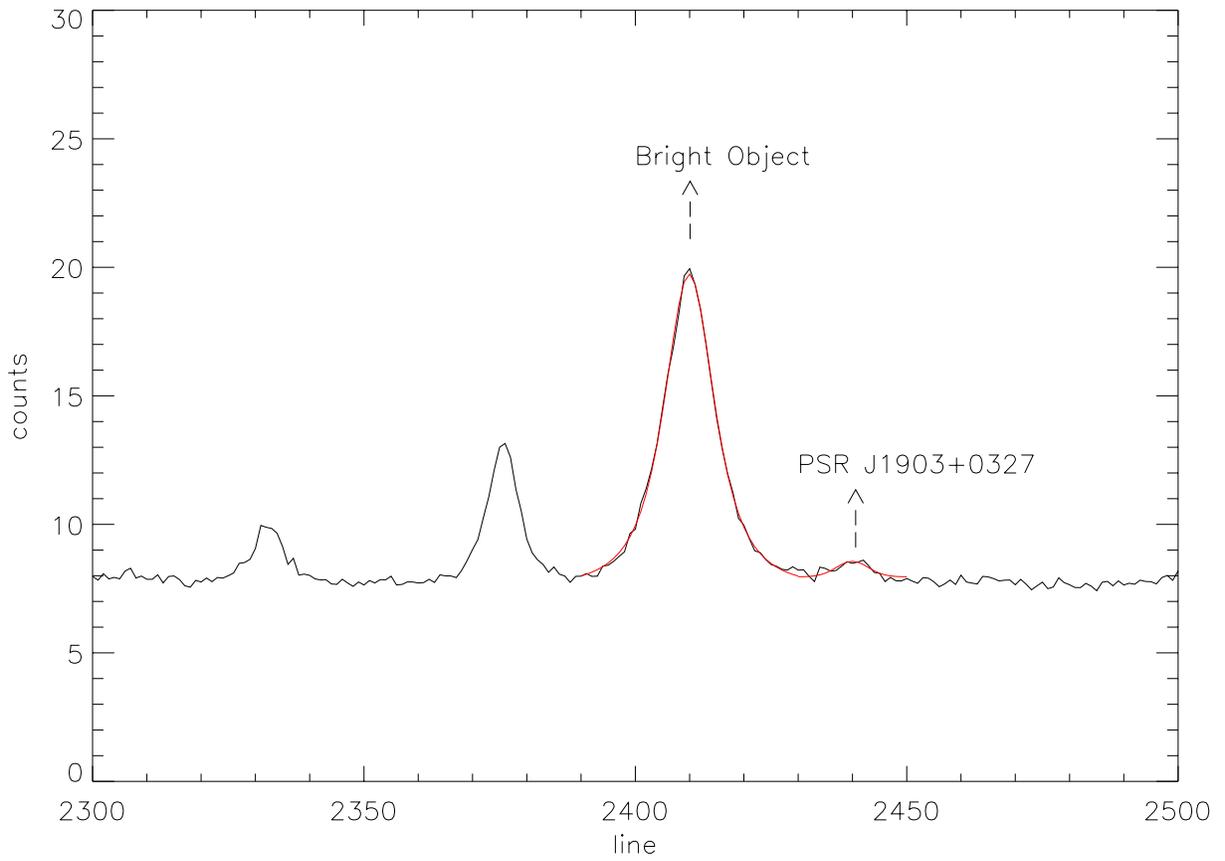}
\caption{The position of the pulsar (line $\#$ 2442) and the bright object (line $\#$ 2410) 2.3$^\prime$$^\prime$ from the pulsar is shown in the above figure for a total of 100 columns. Moffat fits to their profiles indicate negligible contamination to the counts from the pulsar companion.}
\end{figure}

\clearpage

\begin{figure}
\figurenum{2}
\epsscale{1.0}
\plotone{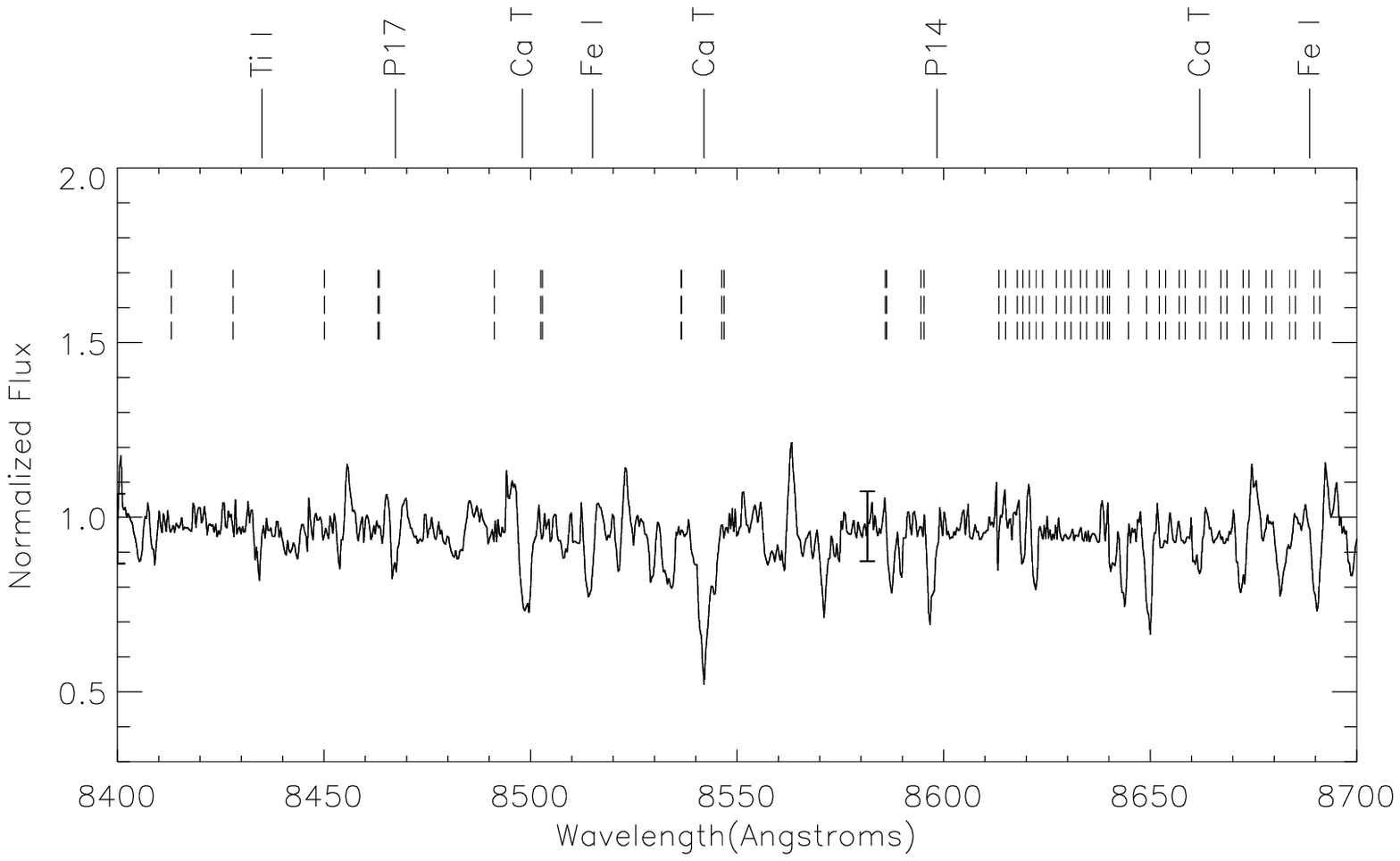}
\caption{The normalized averaged spectrum of the pulsar companion is shown with representative error bars. \ion{Ca}{2} ($\lambda8498$ and $\lambda8542$) lines are visible in the spectrum as well as the P14 line at 8598\AA \ and \ion{Fe}{1} at 8515\AA. There are also possible detections of \ion{Ti}{1} (8435 \AA),\  P17 (8467 \AA)\ and \ion{Fe}{1} (8688.5 \AA)\ lines but we cannot be certain due to the low S/N of the spectrum. The position of the sky lines are indicated by the vertical dashed lines. Sky lines that are too close in wavelengths to be resolved as separate lines, appear as thick dark vertical lines}
\end{figure}

\begin{figure}
\begin{center}
\figurenum{3}
\includegraphics[angle=0,scale=0.50]{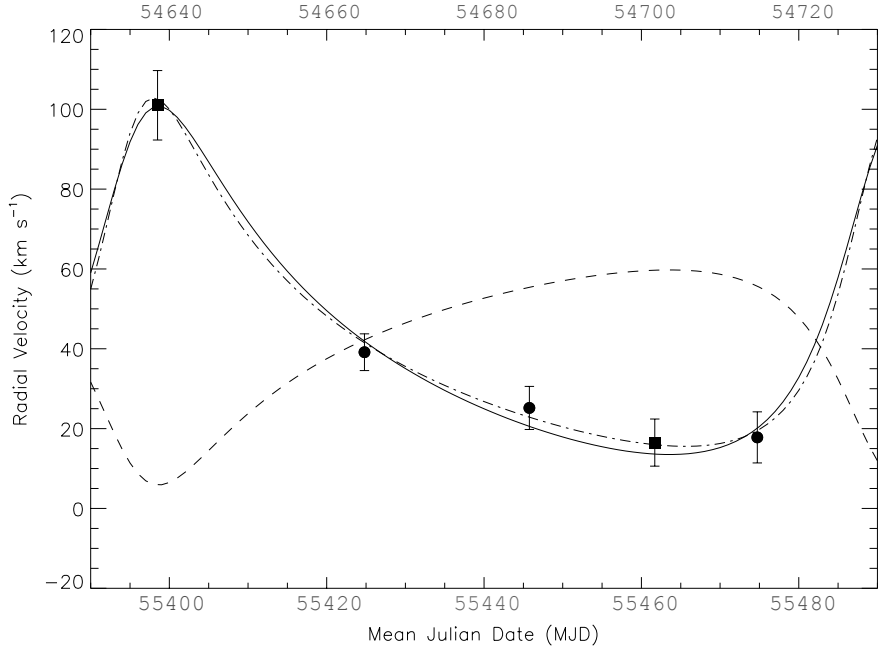}
\caption{The barycentric radial velocity curve of the pulsar (dashed line) and the pulsar companion (solid line), as predicted from pulsar timing ($e=0.44$, $R=1.62$) with $\gamma$=42.1km s$^{-1}$(obtained from optical data). The bottom x-axis and the top x-axis are the Mean Julian Date (MJD) of observations conducted by us and F2011 respectively. The data points in filled circles are the measured values of radial velocity obtained from our dataset while the two points in filled squares (near MJD 54640 and MJD 54700) from F2011 are also shown in the figure. The dotted-dashed line represents the radial velocity curve obtained for $e=0.50$, $R=1.62$ and $\gamma=42.1$ km s$^{-1}$.}
\end{center}
\end{figure}

\clearpage

\begin{figure}
\figurenum{4}
\epsscale{1.0}
\plotone{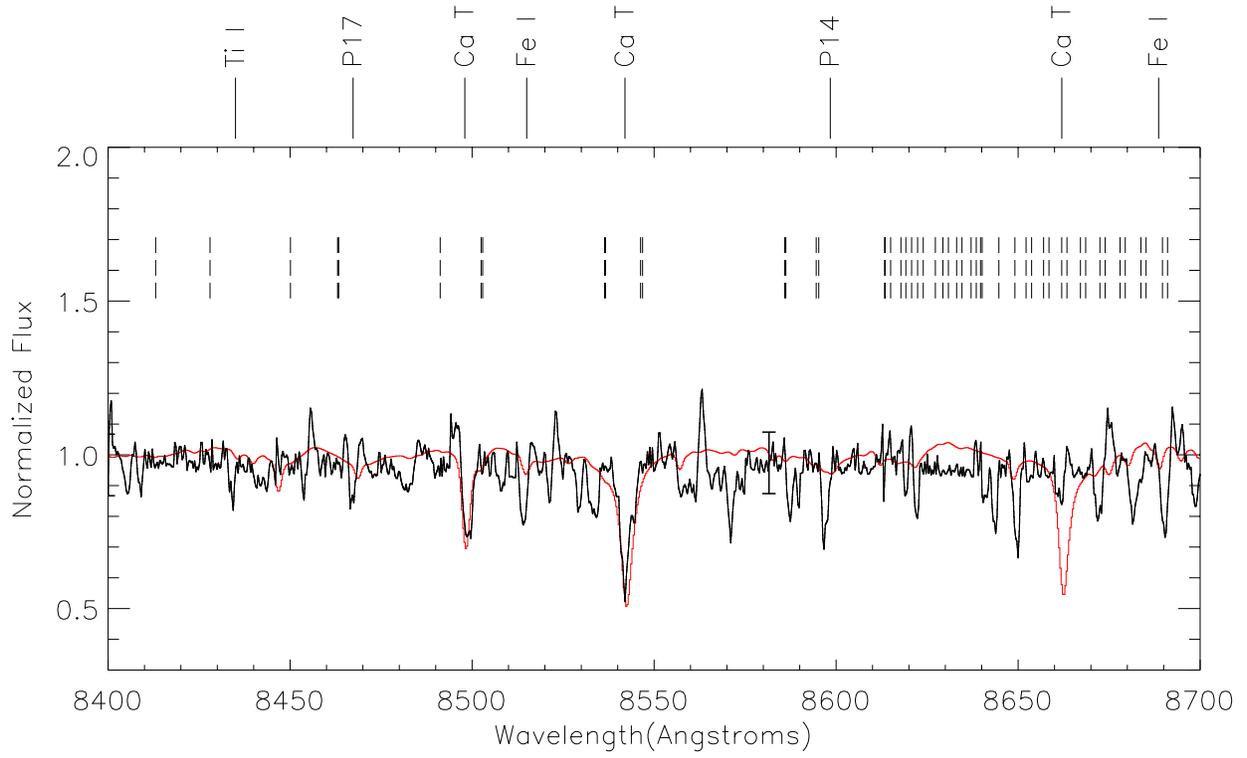}
\caption{The normalized and averaged spectrum of the pulsar companion is shown in the figure compared to the normalized rotational velocity standard star spectrum (in red). The location of the sky lines are marked in vertical dashed lines.}
\end{figure}

\end{document}